\documentstyle[aps,prl,epsf]{revtex}
\begin{document}

\draft  % Makes pacs numbers print (REVTEX)

\wideabs{
\title{Spin liquid ground state in a two dimensional non-frustrated spin model}
\author{G. Santoro$^{1}$, S. Sorella$^{1}$, L. Guidoni$^{1}$, 
A. Parola$^{2}$, and E. Tosatti$^{1,3}$}
\address{$^{(1)}$ Istituto Nazionale per la Fisica della Materia, and
S.I.S.S.A., Via Beirut 2, Trieste, Italy\\
$^{(2)}$ Istituto Nazionale per la Fisica della Materia, and
Universit\`a di Milano, via Lucini 3, Como, Italy\\
$^{(3)}$ International Centre for Theoretical Physics, Strada Costiera,
Trieste, Italy}
%
%%%%%%%%%%%%%%%%%%%%%%%%%%%%%%%%%%%%%%%%%%%%%%%%%%%%%%%%%%%%%%%%%%%%%%%
\date{\today}
\maketitle
\begin{abstract}
We consider an exchange model
describing two isotropic spin-1/2 Heisenberg antiferromagnets coupled by a 
quartic term on the square lattice.
The model is relevant for systems with orbital degeneracy and strong 
electron-vibron coupling in the large Hubbard repulsion limit, and
is known to show a spin-Peierls-like dimerization in one dimension.
In two dimensions we calculate energy gaps, susceptibilities, and correlation 
functions with a Green's Function Monte Carlo. 
We find a finite spin gap and no evidence of any kind of order. 
We conclude that the ground state is, most likely, a spin liquid of 
resonating valence bonds. 
\end{abstract}
 % (REVTEX)
%
%%%%%%%%%%%%%%%%%%%%%%%%%%%%%%%%%%%%%%%%%%%%%%%%%%%%%%%%%%%%%%%%%%%%%%%%%%
%                               TEXT
%%%%%%%%%%%%%%%%%%%%%%%%%%%%%%%%%%%%%%%%%%%%%%%%%%%%%%%%%%%%%%%%%%%%%%%%%%
}		 %end wideabs

%-------------------------------------------------------------------------
%------- Paragrafo in cui scrivo l'hamiltoniana, giustificando la fisica.

The existence of a homogeneous resonating valence bond (RVB) ground state (GS)
for two-dimensional (2D) spin-1/2 systems has been a subject of intense 
theoretical study in the past two decades. 
The idea that the reduced dimensionality and the small spin value might 
enhance quantum fluctuations up to the point of destroying the classical 
N\'eel antiferromagnetic order was first put forward by Anderson in 
1973.\cite{Anderson_73}

The subject became a hot topic after Anderson suggested that the 
parent compounds of the high-$T_c$ copper-oxides
-- argued to be well described by a S=1/2 Heisenberg model on a
square lattice -- might have a spin liquid GS, 
and that superconductivity would result from doping such a spin liquid. 

Since then, many studies on the S=1/2 2D Heisenberg model have shown that 
reduced dimensionality and low spin are not sufficient to 
stabilize a RVB GS. 
First, it is now well established that the square lattice case is 
N\'eel ordered.\cite{reger_young} 
More surprisingly, three-sublattice N\'eel order is also likely to survive 
on the triangular lattice, where the model is frustrated.\cite{bernu,capriotti}
A stronger lattice frustration -- like in the kagom\'e lattice\cite{kagome}) --,
multiple-spin exchange terms on the triangular lattice\cite{misguich}, 
or frustration due to the coupling constants 
-- like in the $J_1-J_2$ Heisenberg model on the square 
lattice\cite{ziman,j1j2} --  
might be more effective in stabilizing a spin liquid GS.
The evidences for a triplet gap and the definite assessment
about the liquid nature of the GS are, however, either limited
to small lattice exact diagonalizations
($N=36$ sites),\cite{kagome,misguich} or must cope, 
when Monte Carlo is used, with the sign problem.\cite{j1j2} 

We propose here a clean example of a RVB ground state realized,
in absence of any frustration, in a 2D model where 
an extra spin-1/2 ${\bf T}$, representing an orbital degree 
of freedom, is coupled to the usual spin ${\bf S}$.
Specifically, the model we have considered is a spin-exchange Hamiltonian 
of the following form:
\begin{equation} \label{hst:eqn}
H_{\rm ST} = -J \sum_{(ij)} X^{(S)}_{i,j} X^{(T)}_{i,j} = 
J \sum_{(ij)} X_{i,j} \;,
\end{equation}
where $J>0$, 
the summation runs over the nearest-neighbor (nn) sites of a square lattice, 
$X^{(S)}_{i,j}=(2 {\bf S}_i \cdot {\bf S}_j-1/2)$, and
$X^{(T)}_{i,j}=(2 {\bf T}_i \cdot {\bf T}_j-1/2)$. 
$H_{\rm ST}$, henceforth referred to as ST model, describes for $J>0$
the low-energy 
physics of an insulating crystal with one electron per site in a two-fold 
degenerate orbital, in the limit of large on-site repulsion (Mott insulator) 
and in presence of Jahn-Teller (JT) effect.\cite{Santoro}
The derivation of Eq.\ (\ref{hst:eqn}), in the same spirit of the derivation 
of the Heisenberg model from the large-U Hubbard limit, 
is standard.\cite{Auerbach,Santoro,Fabrizio} 
The crucial physical condition to be verified is that, among the possible 
two-particle states obtained upon virtual hopping, the {\em inter-orbital 
singlet\/} should turn out to be the lowest in energy, 
which is indeed the case when a strong dynamical JT effect is at play. 
A different, perhaps more common, physical situation is that the lowest
two-particle intermediate state is a Hund's triplet,
in which case the exchange model leads to spin 
ferromagnetism.\cite{KK,Auerbach}
In the general case, the exchange Hamiltonian contains pseudo-spin anisotropic
terms, and some aspects of its phase diagram have been addressed with 
different techniques in Refs.\ \cite{Auerbach,Last_Khomskii}. 

A class of SU(n)-invariant generalizations of the Heisenberg model has been 
previously proposed as candidate non-N\'eel 
antiferromagnets.\cite{Affleck,Read_Sachdev} 
It turns out that the ST model is unitarily equivalent to a particular 
SU(4) model in that class, with $n_c=m=1$ 
(notations of Ref.\ \cite{Read_Sachdev}). 
The $n\to\infty$ limit of these SU(n) models has been
studied in detail, and shown to have, for $n_c=m=1$, a 
spin-Peierls GS both in 2D,\cite{Read_Sachdev,Sachdev,Leung} as well as in one 
dimension (1D).\cite{Affleck}
Recently, we have verified that the spin-Peierls nature of the GS persists,
in 1D, down to the ST model point, i.e., $n=4$.\cite{Santoro}
The behavior in 2D is more subtle. 
We show in the present letter that a RVB GS is found for $n=4$ in 2D, 
in marked contrast with the $n\to \infty$ limit predictions. 

The relationship between the ST model and the usual Heisenberg model -- 
$H_{\rm Heis}=(J/2)\sum_{(ij)}X^{(S)}_{i,j}$ -- is made more
clear by working in the singlet subspace with the overcomplete 
basis set of the valence bond (VB) configurations, i.e., products of 
non-overlapping singlet dimers $(i,j)$ 
connecting pairs of sites on opposite sublattices and covering the lattice. 
More precisely, if
$(i,j)_{S(T)}=(\uparrow_i\downarrow_j-\downarrow_i\uparrow_j)_{S(T)}$ 
denotes a singlet bond between sites $i$ and $j$ for the S (T) variables, 
we restrict the VB configurations considered to the invariant subspace 
-- containing the GS, by positivity arguments -- 
in which every singlet dimer $(i,j)$ is a product of
an S- and a T-singlet, $(i,j)=(i,j)_S (i,j)_T$. 
Since $(i,j)_{S(T)}=-(j,i)_{S(T)}$, we assume the sign 
convention that the leftmost index in the singlet pair always belongs 
to sublattice A. 
It is possible to show that the GS of both the ST and the 
Heisenberg model can be written as linear combinations with positive 
coefficients of VB configurations,
with the previous restrictions and sign conventions.
The construction of excellent variational wavefunctions based on combinations
of VB configurations has been shown to be possible even for the Heisenberg
case provided sufficiently long-ranged bonds are allowed.\cite{Doucot}
On the other hand, it is quite clear that whenever short-ranged bonds are
the dominant ones, the GS will have a characteristic length $\xi$
and there will be a finite gap, in the thermodynamic limit, to the lowest
triplet excitations.  
 
The bond operators 
$X^{(S)}_{i,j}=(2 {\bf S}_i \cdot {\bf S}_{j}-1/2)$, and
similarly $X^{(T)}_{i,j}$, have
simple properties when acting on VB configurations, because they affect
at most two singlet pairs.  
Indeed, it is known that\cite{Anderson_73} 
$X^{(S)}_{i,j} (i,j)_S = -2(i,j)_S$, and 
$X^{(S)}_{j,k} (i,j)_S (k,l)_S = -(k,j)_S (i,l)_S$. 
Using these rules for $X^{(S)}_{i,j}$ and $X^{(T)}_{i,j}$,
it is straightforward to show that the bond operators
$X_{i,j}=-X^{(S)}_{i,j} X^{(T)}_{i,j}$ appearing in the ST model,
Eq.\ (\ref{hst:eqn}), obey the following rules:
\begin{eqnarray} \label{chi_rules:eqn}
X_{i,j} (i,j) &=& -n \; (i,j) \;, \nonumber \\
X_{j,k} (i,j) (k,l) &=& -(k,j) (i,l)  \;, 
\end{eqnarray}
with $n=4$. 
Notice that these relationships are formally identical to those relevant to 
the Heisenberg case, except for a coefficient $n=4$, in place of $n=2$, 
when a nn Hamiltonian bond $(ij)$ acts on a single dimer $(i,j)$. 
This enhanced coefficient favors the formation of short-ranged bonds in the
GS of the ST model, making the suppression of N\'eel long range
order (LRO) more likely. 
% QUI: Aggiustare 
In the limit $n\to \infty$ we recover, from Eq.\ (\ref{chi_rules:eqn}), 
the known results both in 1D,\cite{Affleck} and 2D.\cite{Read_Sachdev}
Indeed, for $n\to\infty$ the only surviving VB configurations are those 
with nn dimers only. 
In 1D this leads to a doubly-degenerate spin-Peierls GS.\cite{Affleck}
In 2D the model maps\cite{Read_Sachdev} onto the purely kinetic limit of the 
quantum dimer model (QDM) of Rokhsar and Kivelson.\cite{Rok_Kiv} 
(This purely kinetic point of the QDM is believed to be 
characterized by a plaquette resonating VB state,\cite{Leung} 
breaking translational invariance.) 

By working with the VB basis, the action of the Hamiltonian $H \psi_i$ on any 
basis element $\psi_i$ defines a non-symmetric matrix, 
$H \psi_i = \sum_j h_{j,i} \psi_j$, with all elements $h_{j,i}$ non-positive,
as implied by Eq.\ (\ref{chi_rules:eqn}). 
The right eigenvector of $h_{j,i}$ with minimum eigenvalue,
corresponding for $h_{j,i}\le 0$ to the GS of $H$, 
can be computed by applying the power method, as implemented stochastically by 
means of the Green's Function Monte Carlo (GFMC) method. 
The GFMC is in fact not limited to symmetric matrices, 
and there is no sign problem when all $h_{j,i}$ are non-positive.\cite{j1j2}
Using the VB basis, the GFMC turns out to have extremely small statistical
errors, compared to the more conventional algorithm\cite{Calandra} employing 
an Ising basis. 
In this formulation, the GFMC does not require the calculation of the
overlaps $\langle \psi_i \mid \psi_j \rangle$ between VB configurations. 
Details of the method are given elsewhere.\cite{new_GFMC} 

This new and simple GFMC allows us to obtain a very accurate 
determination of the triplet gap 
by performing two independent simulations of the singlet GS 
and the triplet lowest excited state. 
In the latter case, the basis employed is slightly
modified, by allowing for the presence of a single triplet bond 
$(i,j)^{t}=(\uparrow_i\downarrow_j+\downarrow_i\uparrow_j)_S (i,j)_T$ 
in each element of the VB basis. The corresponding rules for the application
of $X_{i,j}$ are: $X_{i,j} (i,j)^{t} = 0$, and 
$X_{j,k} (i,j)^t (k,l) = -(k,j) (i,l)^t$. 
Notice that this implies the absence of sign problem in the triplet 
subspace as well. 

\begin{figure} 
\centerline{\epsfxsize=3.0in\epsfysize=2.6in\epsfbox{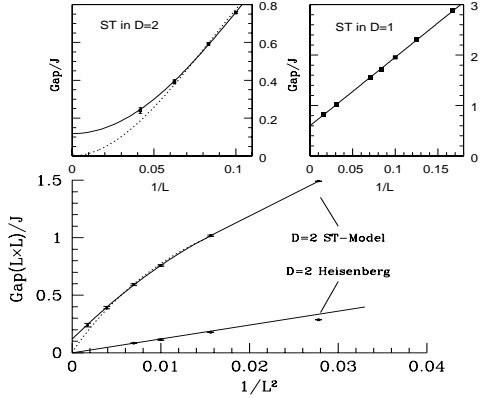}}
\vspace{-1.0cm}
\caption{
Finite size gap between the GS and the first excited
triplet state for the ST model and for the Heisenberg
model\protect\cite{Calandra} on the square lattice. 
The ST data are obtained by the new GFMC method working with the VB basis. 
The solid and dashed lines are different fits described in the text. 
The data for the ST model clearly indicate the existence of
a spin gap in the thermodynamic limit. 
Left inset: The ST data plotted versus $1/L$. 
Right inset: The triplet gap for the ST in 1D. 
}
\label{gap2d:fig}
\end{figure}

%------------------------------------------------------------------------
%--------------------------- 2D RESULTS ---------------------------------
Fig.\ \ref{gap2d:fig} shows the results obtained for the triplet gap 
for $L\times L$ square lattices with $L$ up to $24$, as a function of the
inverse volume. 
The corresponding data for the Heisenberg case,\cite{Calandra} also shown 
for comparison, are consistent with a finite size gap $\Delta_L$ 
scaling to zero as $a/L^2+b/L^3+\cdots$. 
The dashed lines (see also left inset in Fig.\ \ref{gap2d:fig}) show our 
best two-parameter fit to the ST data obtained by imposing the same gap 
behavior as in the Heisenberg case: such a fit is clearly unsatisfactory. 
Instead, the solid line through the ST data is the result of a three parameter
fit of the form $\Delta_L=\Delta+a/L^2+b/L^4+\cdots$,\cite{gap_scaling} 
giving a clear evidence of a finite gap $\Delta$ in the thermodynamic limit.   
Clearly, the detailed finite-size behavior of the ST gap in 2D 
is non trivial, requiring the simulation of quite large lattices ($L=24$) to 
pin-down the presence of a gap. 
This suggests the presence of a length scale $\xi$ of
the order of $10\div 20$ lattice spacings.
This behavior should be contrasted to that of the triplet gap for the 
ST model in 1D, shown in the right inset of Fig.\ \ref{gap2d:fig}, where the 
size scaling of the gap is straightforward ($\xi \approx 1$).

\begin{figure} 
\centerline{\epsfxsize=3.0in\epsfysize=2.6in\epsfbox{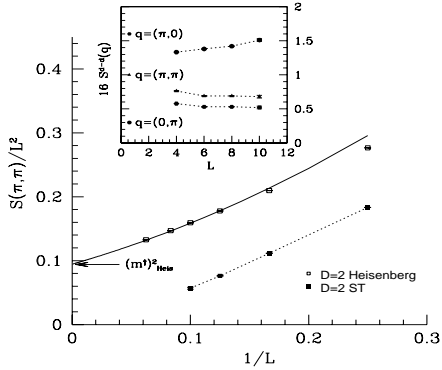}}
\vspace{-1.0cm}
\caption{Finite size structure factors obtained by GFMC with the forward 
walking technique. 
From the spin-spin $S(\pi,\pi)$ we find no clear sign of N\'eel LRO. 
The size scaling of the N\'eel order parameter for the 
Heisenberg model\protect\cite{Calandra} is shown for comparison.
Inset: The important components of the Fourier transform of
the dimer-dimer correlation functions appear to be finite. 
}
\label{correl:fig}
\end{figure}

In principle, either a VB crystal with some broken spatial symmetry, as in 1D, 
or a homogeneous spin liquid is compatible with the existence of a spin gap. 
In order to investigate the possible kinds of LRO
which might characterize the GS of the ST model in 2D, 
we have calculated the expectation value of several spin-spin
correlation functions by the forward walking technique.\cite{Runge,Calandra}
Fig.\ \ref{correl:fig} shows the results for the spin structure factor
$S(\pi,\pi)=\sum_{\bf r} (-1)^{\bf r} 
\langle {\bf S}_0 \cdot {\bf S}_{\bf r} \rangle$, which should diverge with
the system volume $L^2$, when the N\'eel order parameter $m^{\dagger}$ is
non-vanishing,  $S(\pi,\pi)\sim (L m^{\dagger})^2$. 
The comparison between the Heisenberg and the ST model results for
$S(\pi,\pi)$ strongly suggests the absence of N\'eel order in the ST case,
in agreement with the presence of a triplet gap. 
More interestingly, the inset in Fig.\ \ref{correl:fig} shows the 
results obtained, in the ST case, for the dimer-dimer structure factors 
with nn bonds in the $\hat{x}$-direction, 
$S^{d-d}({\bf q})=\sum_{\bf r} e^{i{\bf q}\cdot {\bf r}} 
\langle S^z_0 S^z_{0+\hat{x}} S^z_{\bf r} S^z_{{\bf r}+\hat{x}} \rangle$. 
The ${\bf q}=(\pi,\pi)$ and ${\bf q}=(0,\pi)$ dimer structure factors are
clearly finite in the $L\to \infty$ limit; the ${\bf q}=(\pi,0)$ also does
not seem to diverge linearly with the volume, but the present data are limited
to too small sizes ($L\le 10$) to be conclusive. 

\begin{figure} 
\centerline{\epsfxsize=3.0in\epsfysize=2.6in\epsfbox{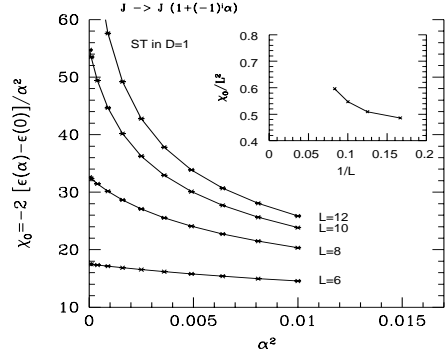}}
\vspace{-1.0cm}
\caption{The susceptibility to breaking translation symmetry for the ST model
in 1D, which is known to have a spin-Peierls-like dimerization.
Data are obtained by exact diagonalizations.
Inset: the extrapolated $\alpha=0$ value of $\chi_O$ divided by the square
of the system volume $L^2$. 
}
\label{trasla1d:fig}
\end{figure}

\begin{figure} 
\centerline{\epsfxsize=3.0in\epsfysize=2.6in\epsfbox{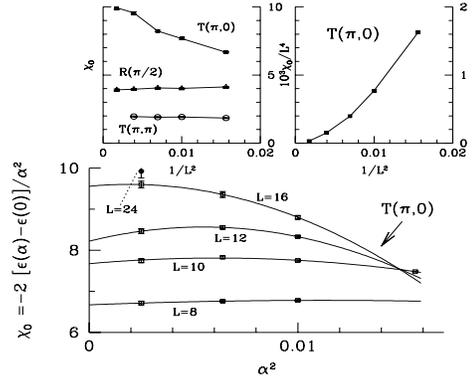}}
\vspace{-1.0cm}
\caption{Susceptibilities to breaking the most important crystal symmetries
for the ST in 2D. 
The finite-size, finite-$\alpha$ data for $\chi_O$ can be readily extrapolated,
by quadratic fits, to $\alpha=0$ values: notice the distinctly different
behavior for the corresponding 1D results of Fig.\ \ref{trasla1d:fig}.
The insets summarize the results obtained for the extrapolated $\chi_O$. 
}
\label{chi2d:fig}
\end{figure}

In order to have further evidence about the existence of a true 
homogeneous spin liquid GS, we have directly calculated, using 
the new more efficient GFMC algorithm described previously, the response of
the system to symmetry-breaking operators.
It is practically impossible to exclude all possible types of crystalline
order numerically; in the following we restrict our consideration to the
most plausible types of order, involving either a broken translation symmetry
$T$ with momenta compatible with a real GS 
(${\bf q}=(\pi,0),(0,\pi),(\pi,\pi)$), or/and a broken $\pi/2$ rotation 
symmetry $R$. This includes all types of crystalline dimer and plaquette order 
proposed so far.\cite{Rok_Kiv,Sachdev,Leung}
More precisely, we have perturbed the ST Hamiltonian by adding a term
$\alpha \hat{O}$, with $\hat{O}$ an operator which breaks one of the symmetries
above: 
$\hat{O}=\sum_{\bf r} e^{i{\bf q}\cdot {\bf r}} X_{{\bf r},{\bf r}+\hat{x}}$
with the appropriate ${\bf q}$ for the translation case, and 
$\hat{O}=\sum_{\bf r}(X_{{\bf r},{\bf r}+\hat{x}}-X_{{\bf r},{\bf r}+\hat{y}})$ 
for the $\pi/2$ rotation. 
On finite size, the GS expectation value of $\hat{O}$ vanishes 
by symmetry, and the GS energy per site has corrections proportional
to $\alpha^2$, 
$\epsilon_{\alpha}=\epsilon_0-\chi_O {\alpha}^2/2$, $\chi_O$ being the 
generalized susceptibility associated to the 
symmetry-breaking operator $\hat{O}$.
On the other hand, if symmetry breaking occurs in the thermodynamic
limit, it is possible to have $\lim_{\alpha\to 0} \lim_{L\to \infty} \langle
\hat{O}/L^2 \rangle = p \ne 0$.
In the latter case, $\chi_O$ has to diverge as $L\to \infty$. 
More precisely, it is possible to show that $\chi_O$ is bounded from below
by the order parameter times the system volume squared,
$\chi_O > {\rm const}\; p^4 L^4$. 
These susceptibilities are therefore a very sensitive tool --
much more than the usual square of the order parameter -- for detecting LRO. 
For instance, as shown in Fig.\ \ref{trasla1d:fig} for the ST model in 1D,
the presence of dimerization in the thermodynamic limit\cite{Santoro} 
is readily inferred from the behavior of $\chi_O$, with 
$\hat{O}=\sum_i (-1)^i X_{i,i+1}$,
even for very small system sizes ($L\le 12$). 
Fig.\ \ref{chi2d:fig} shows the results obtained for the ST model in 2D.
For all the symmetry breaking operators considered, the associated 
susceptibilities are found to be finite. 
The largest susceptibility is found to be the one associated to 
columnar order ($T(\pi,0)$ in Fig.\ \ref{chi2d:fig}), and is only 
weakly increasing with size, eventually saturating to a constant, 
in marked contrast to the strong divergence of the 1D analog 
(see Fig.\ \ref{trasla1d:fig}).  

%--------------------- CONCLUSIONS: ---------------------------------------
In conclusion, we have studied a non-frustrated exchange 
Hamiltonian, Eq.\ (\ref{hst:eqn}), 
which describes the low-energy physics of a Mott insulator with orbital
degeneracy in the regime in which the inter-orbital singlet is the 
lowest-energy intermediate state available to virtual hopping. 
We find a clear evidence for a spin gap without crystalline VB order in 2D.
A homogeneous liquid of resonating valence bonds appears as the natural 
candidate GS for this model. 
An important question about the nature of the low-lying excitations 
remains to be addressed: is there a branch of gapless excitations in 
the singlet sector, or the system is gapped to all excitations?
The answer to this question will require future work. 

ACKNOWLEDGMENTS --
We acknowledge partial support from INFM through PRA HTSC and LOTUS,
from MURST through COFIN97, and from EU, through contract 
FULPROP ERBFMRXCT970155.

%%%%%%%%%%%%%%%%%%%%%%%%%%%%%%%%%%%%%%%%%%%%%%%%%%%%%%%%%%%%%%%%%%%%%%%%%
%                               BIBLIOGRAPHY
%%%%%%%%%%%%%%%%%%%%%%%%%%%%%%%%%%%%%%%%%%%%%%%%%%%%%%%%%%%%%%%%%%%%%%%%%

%%%%%%%%%%%%%%%%%%%%%%%%%%%%%%%%%%%%%%%%%%%%%%%%%%%%%%%%%%%%%%%%%%%%%%%%%
%                               FIGURES
%%%%%%%%%%%%%%%%%%%%%%%%%%%%%%%%%%%%%%%%%%%%%%%%%%%%%%%%%%%%%%%%%%%%%%%%%
%
%------------------------------------------------------------------------
%------------------------------------------------------------------------
%
%------------------------------------------------------------------------
%------------------------------------------------------------------------
%
%------------------------------------------------------------------------
%------------------------------------------------------------------------
%
%------------------------------------------------------------------------
%------------------------------------------------------------------------


\begin{thebibliography}{99}

\bibitem{Anderson_73}
P. W. Anderson, Mater.\ Res.\ Bull.\ {\bf 8}, 153 (1973).

\bibitem{reger_young}
J. D. Reger and A. P. Young, Phys.\ Rev.\ B {\bf 37}, 5978 (1988).

\bibitem{bernu}
B. Bernu, C. Lhuillier, and L. Pierre,
Phys.\ Rev.\ Lett.\ {\bf 69}, 2590 (1992).

\bibitem{capriotti} 
L. Capriotti, A. Trumper, and S. Sorella,
preprint cond-mat/9901068.

\bibitem{kagome}
C. Waldtmann, H.U. Everts, B. Bernu, C. Lhuillier, P. Sindzingre,
P. Lecheminant, L. Pierre, Eur.\ Phys.\ J. B, {\bf 2}, 501 (1998). 

\bibitem{misguich}
G. Misguich, B. Bernu, C. Lhuillier, C. Waldtmann, Phys.\ Rev.\ Lett.\
{\bf 81}, 1098 (1998). 

\bibitem{ziman}
H. Schulz and T. Ziman, Europhys.\ Lett.\ {\bf 18}, 355 (1992). 

\bibitem{j1j2}
S. Sorella, Phys.\ Rev.\ Lett.\ {\bf 80}, 4558 (1998).

\bibitem{Santoro}
G. Santoro, L. Guidoni, A. Parola, and E. Tosatti, 
\prb {\bf 55}, 16168 (1997);  
L. Guidoni, G. Santoro, S. Sorella, A. Parola, and E. Tosatti, 
J. Appl.\ Phys.\ (1999). 

\bibitem{Auerbach}
D. P. Arovas and A. Auerbach, Phys.\ Rev.\ B {\bf 52}, 10114 (1995).

\bibitem{Fabrizio}
M. Fabrizio, M. Airoldi, and E. Tosatti, Phys.\ Rev.\ B {\bf 53}, 12086 (1996).

\bibitem{KK}
For a review see, for instance, 
K. I. Kugel' and D. I. Khomskii, Sov.\ Phys.\ Usp.\ {\bf 25}, 231 (1982).

\bibitem{Last_Khomskii} 
S. K. Pati, R. R. P. Singh, D. I. Khomskii,
Phys.\ Rev.\ Lett.\ {\bf 81}, 5406 (1998).

\bibitem{Affleck}
I. Affleck, Phys.\ Rev.\ Lett.\ {\bf 54}, 966 (1985).

\bibitem{Read_Sachdev}
N. Read and S. Sachdev, Nucl.\ Phys.\ B{\bf 316}, 609 (1989).

\bibitem{Sachdev}
S. Sachdev, Phys.\ Rev.\ B {\bf 40}, 5204 (1989).

\bibitem{Leung}
P. W. Leung, K. C. Chiu, and K. J. Runge,  Phys.\ Rev.\ B {54}, 12938 (1996). 

\bibitem{Doucot}
S. Liang, B. Doucot, and P. W. Anderson, \prl
{\bf 61}, 365 (1988).

\bibitem{Rok_Kiv}
D. Rokhsar and S. Kivelson, \prl {\bf 61}, 2376 (1988).

\bibitem{Calandra}
M. Calandra and S. Sorella, Phys Rev. B {\bf 57}, 11446 (1998).

\bibitem{new_GFMC}
S. Sorella and G. Santoro (unpublished). 

\bibitem{gap_scaling}
This form is consistent with low-lying triplet excitations of the form
$\sqrt{\Delta^2+c^2k^2}$, with $k\approx 2\pi/L$, dominating the finite-size 
corrections over exponentially small terms in the scaling of the
GS energy. 

\bibitem{Runge}
K. Runge,  Phys.\ Rev.\ B {\bf 45}, 7229 (1992); {\it ibid.\/} {\bf 45}, 
12292 (1992).

\end{thebibliography}
\end{document}